\documentclass[pra,twocolumn,amsmath,amssymb,groupedaddress]{revtex4}
\usepackage{graphicx}

\def\be{\begin{equation}}
  \def\ee{\end{equation}}

\newcommand{\K}{{\cal K}}
\newcommand{\M}{{\cal M}}
\newcommand{\Tr}{\textmd{Tr}}

\begin{document}

\title{Effective Spin Quantum Phases in Systems of Trapped Ions}

\author{X.-L. \surname{Deng}}
\email{Xiaolong.Deng@mpq.mpg.de}
\author{D. \surname{Porras}}
\email{Diego.Porras@mpq.mpg.de}
\author{J.~I. \surname{Cirac}}
\email{Ignacio.Cirac@mpq.mpg.de}
\affiliation{Max-Planck-Institut f\"ur Quantenoptik, Hans-Kopfermann-Str. 1, Garching, D-85748, Germany.}

\begin{abstract}
A system of trapped ions under the action of off--resonant standing--waves can be used to simulate a variety of quantum spin models. In this work, we describe theoretically quantum phases that can be observed in the simplest realization of this idea: quantum Ising and XY models. Our numerical calculations with the Density Matrix Renormalization Group method show that experiments with ion traps should allow one to access general properties of quantum critical systems. On the other hand, ion trap quantum spin models show a few novel features due to the peculiarities of induced effective spin--spin interactions which lead to interesting effects like long--range quantum correlations and the coexistence of different spin phases.
\end{abstract}

\date{\today}

\maketitle

\section{Introduction}
The goal of quantum many--body physics is the understanding of properties of materials ranging from quantum magnets to high--Tc superconductors. For this task, theorists have developed a variety of simplified models which allow us to describe the rich phenomenology that is observed in those systems.
In the last years, progress in the manipulation of matter at the microscopic scale has changed dramatically this situation.
Atomic experimental systems such as cold atoms in optical lattices offer us the possibility to tune interactions and engineer at will quantum phases. In this way quantum interacting models can be studied in a controlled way and some of the limitations of solid-state set-ups are overcome.
A successful step in this direction was given recently with the observation \cite{mott} of the superfluid--Mott insulator transition  with bosonic atoms in an optical lattice. Recently, cold atoms in optical lattices have also been proposed for the simulation of certain quantum spin models \cite{optical.lattices}.

On the other hand trapped ions are a clean experimental system where quantum optics offers us accurate techniques for the  manipulation and measurement of quantum states \cite{Leibfried.review}.
In the last years this experimental field has been mainly motivated by applications to quantum information processing \cite{NielsenChuang}, in which internal electronic states are used as qubits, and vibrational modes permit us to perform quantum gates between them \cite{CiracZoller95,ion.QC}.
In fact, by using similar ideas we have shown recently that trapped ions can be used for the study of a rich variety of quantum interacting models \cite{Spin.Simulator,BHM.Phonons}. The experimental requirements for the study of quantum interacting systems with trapped ions are indeed much less stringent than those for quantum information tasks. 

In this paper we pursue the ideas proposed in Ref. \cite{Spin.Simulator} for the realization of a quantum simulator \cite{quantum.simulation,Jane} of quantum magnetism with trapped ions. We focus on the possibility of studying antiferromagnetic Ising and XY spin chains in linear ion traps, where quantum phase transitions can be induced and explored \cite{Sachdev}. We show that experiments with trapped ions can access the physics of magnetic quantum phases with an accuracy that is not possible in other experimental set-ups, and allow us to test general properties of quantum systems at, or near, criticality. On the other hand, the effective spin Hamiltonians that can be engineered with trapped ions show a few remarkable features, the most important one being the fact that spin--spin interactions follow power--law decays, something that induces long--range quantum correlations between distant ions, which are absent in the usual nearest--neighbor models.

Our paper is organized as follows. 
In the first section we describe in detail our proposal, taking as a starting point the system formed by a set of trapped ions coupled to an off--resonant standing wave, and show that it realizes a system of effective interacting spins. We focus on the cases of Ising and XY interactions. 
In the second section we study the many--body problem posed by the quantum Ising model in ion traps. For this purpose, we make use of a quasi--exact numerical method, the so called Density Matrix Renormalization Group \cite{White}, and introduce also an analytical, approximate solution in terms of Holstein--Primakoff bosons. The third section deals with the study of the XY model in ion traps. Finally, we present our conclusions.

\section{Effective spin quantum systems}

In this section we review the theoretical description of a set of trapped ions under the action of off--resonant standing waves presented in \cite{Spin.Simulator}, and show that their quantum dynamics follows approximately that of quantum interacting spins. We also discuss carefully the effects of residual spin--phonon couplings.

Our proposal relies on an always-on coupling between internal states and vibrational modes, in such a way that phonons transmit an interaction between different ions. This paper deals, thus, with the theoretical description of a system of effective spins (internal states) coupled to a set of vibrational modes. The corresponding Hamiltonian has three contributions:
\begin{equation}
H = H_m + H_v + H_f .
\label{Hamiltonian.initial}
\end{equation}
$H_m$ describes the local dynamics of internal states, which are a set of independent effective spins under the action of effective global magnetic fields in each direction:
\begin{equation}
H_m = \sum_{\substack{j=1, \ldots, N \\ \alpha = x,y,z}} B^\alpha \sigma_j^\alpha .
\label{Hm}
\end{equation}
Note that $B^z$ is the energy of the internal state, and $B^x$, $B^y$ can be implemented by the action of lasers resonant with the internal transition. In Eq. (\ref{Hamiltonian.initial}), $H_v$ is the vibrational Hamiltonian, and $H_f$ is the internal state--phonon coupling. Both terms are described in the following two subsections.

\subsection{Vibrational modes of ion chains}
We consider 1D systems of trapped ions, whose physical implementation corresponds to Coulomb chains in linear Paul traps or linear arrays of ion microtraps.
Let us assign $\bf{z}$ to the axis of the ion chain, and $\bf{x}$, $\bf{y}$ to the radial directions.
The potential experienced by the ions is determined by the trapping frequencies, $\omega_{\alpha}$ (in the following the index $\alpha$ always runs over the spatial directions, $\alpha = x,y,z$), and the Coulomb repulsion:
\begin{eqnarray}
V &=&
\frac{1}{2} m \sum_j
\left( \omega_x^2 x_j^2 + \omega_y^2 y_j^2  + \omega_z^2 z_j^2 \right) +
\nonumber \\
&& \sum_{j>i} \frac{e^2}{\sqrt{(x_i-x_j)^2 + (y_i-y_j)^2 + (z_i-z_j)^2}} .
\label{vibrational}
\end{eqnarray}
In the harmonic approximation we express $V$ as a function of the displacements around the equilibrium positions ($q^\alpha_j = x^\alpha_j - x^{\alpha,0}_j$), up to quadratic terms:
\begin{eqnarray}
V &=& \frac{1}{2} m \sum_{\alpha,i,j} {\cal K}^{\alpha}_{i,j} {q}^{\alpha}_{i} {q}^{\alpha}_{j} . \nonumber \\
{\cal K}^{\alpha}_{i,j}
&=& \left\{ \begin{array}{ll} \omega^2_{\alpha} - c_{\alpha} \sum_{j' (\neq i)} \frac{e^2 / m }{|z_i^{0} - z^{0}_{j'}|^3} \hspace{0.5cm} i=j
                 \\  \hspace{0.5cm} + c_{\alpha} \frac{e^2 / m }{|z_i^{0} - z_j^{0}|^3} \hspace{0.5cm} i \neq j \end{array} . \right. 
\label{motion}
\end{eqnarray}
The corresponding Hamiltonian can be diagonalized in terms of collective modes (phonons):
\begin{eqnarray}
H_v = \frac{1}{2} m \sum_{i,j,\alpha} {\K_{i,j}^\alpha} q_i^{\alpha} q_j^{\alpha} + \! \! \sum_{i, \alpha} \frac{p^{\alpha}_j}{2 m} = \sum_n \hbar \omega_{\alpha, n} a^{\dagger}_{\alpha,n} a_{\alpha,n},
\end{eqnarray}
where
$p_i^{\alpha}$ are the momenta corresponding to $q^{\alpha}_i$. Local coordinates can be expressed in terms of creation and annihilation of collective vibrational modes:
\begin{equation}
q^{\alpha}_i = \sum_n \frac{\M^{\alpha}_{i,n}}{\sqrt{2 m \omega_{\alpha,n}/\hbar}} \left(a^{\dagger}_{\alpha,n} + a_{\alpha,n} \right) ,
\label{collective.modes}
\end{equation}
where $c_{x,y} = 1$, $c_z = - 2$. Matrices $\cal M^{\alpha}$ in (\ref{conditional.force}) diagonalize the vibrational hamiltonian: $\sum_{i,j} {\cal M}^{\alpha}_{i,n}{\cal K}^{\alpha}_{i,j} {\cal M}^{\alpha}_{j,m} = \omega_{\alpha,n}^2 \delta_{n,m}$.

The characteristics of the vibrational modes are governed by the following parameters, which quantify the relative value of Coulomb interaction and trapping potentials:
\begin{equation}
\beta_{\alpha} = |c_{\alpha}| e^2 / m \omega_{\alpha}^2 d_0^3 ,
\label{beta.ratios}
\end{equation}
where $d_0$ is the mean distance between ions. If $\beta_{\alpha} \ll 1$, then phonons are close to be localized at each ion (stiff limit); on the contrary $\beta_{\alpha} \gg 1$ (soft limit) implies that Coulomb repulsion prevails over the trapping potential and, thus, phonons have a strong collective character that results in the ability to mediate interactions with a range of the order of the ion trap \cite{Spin.Simulator}.

We summarize below the description of vibrational modes in different trapping conditions:
\begin{itemize}
\item[(i)] {\it Coulomb chains (Paul traps).}
In this experimental set--up the equilibrium positions in the axial ($\bf z$) direction are determined by the Coulomb interaction and the axial trapping, and $\beta_z$ depends only on the number of ions $N$. This dependence can be estimated in the limit of many ions (see \cite{Dubin}):
\begin{equation}
\beta_z \approx \frac{1}{12} \frac{N^2}{\log(6 N)}, \hspace{.5cm} N \gg 1 .
\label{beta.par}
\end{equation}
$\beta_z \gg 1$ and thus axial modes in Coulomb crystals are always in the soft limit. On the other hand, $\beta_{x,y}$, can be reduced at will because one can increase the axial trapping frequencies $\omega_{x,y}$ while leaving constant the mean distance between ions, $d_0$. Indeed, condition $\beta_{x,y} < 1$ has to be fulfilled to make the Coulomb chain stable against zig-zag instabilities \cite{Dubin}. 
\item[(ii)] {\it Linear arrays of ion traps.}
It is worth considering this case here due to the experimental effort currently being devoted to the fabrication of linear arrays of ion traps where equilibrium positions of the ions are chosen by individual confinement potentials for each ion \cite{arrays}. In this case, all trapping frequencies $\omega_\alpha$ can be chosen at will, and in particular condition $\beta_z \ll 1$ can be also fulfilled opposite to the case of standard Paul traps.
\end{itemize}

\subsection{Internal state conditional forces}
Internal states are coupled to the motion by placing the ions in an off--resonant standing--wave, such that they experience a state--dependent a.c.--Stark shift. Several configurations are possible, in which internal states are coupled to vibrational modes that are transverse or longitudinal with respect to the trap axis.
The characteristics of the resulting effective spin--spin interaction depends on the choice of the directions of the laser beams: in particular, axial vibrational modes mediate long--range interactions, with a range of the order of the ion chain, and radial modes mediate shorter range spin--spin interactions, with a power--law decay $J_{i,j} \propto 1/|i-j|^3$. Due to the fact that short--range spin Hamiltonians contain a rich quantum critical phenomenology, we focus in this work on this last situation.

Let us consider the following coupling between effective spins and radial motion:
\begin{equation}
H_f = -F_x \sum_j x_j | 1 \rangle \langle 1 |_{z,j} - F_y \sum_j y_j | 1 \rangle \langle 1 |_{y,j} .
\label{conditional.force.0}
\end{equation}
$| 1 \rangle_\alpha$ is the eigenstate of $\sigma^\alpha$ with eigenvalue $1$. The reason
for this choice of couplings is that conditional forces are most
easily implemented with the z component of the internal states
(one only needs a single standing--wave). Thus, it is advantageous
to couple $\sigma^z$ to one of the radial directions. The first
term in $H_f$ is the one which corresponds to the pushing gate
presented in \cite{Cirac00}, while the second one can be
implemented by using two additional standing--waves
\cite{Spin.Simulator}.

The coupling of the internal states to the motion can be written as an effective spin--phonon coupling by expressing the ions' coordinates in terms of collective modes:
\begin{equation}
H_{f} = - \sum_{\alpha, i, n}
F_{\alpha} \frac{{\cal M}^{\alpha}_{i,n}}{\sqrt{2 m \omega_{\alpha,n} / \hbar}}
\left( a^{\dagger}_{\alpha,n} + a_{\alpha,n} \right) \left(1 + \tilde{\sigma}_i^{\alpha} \right) .
\label{conditional.force}
\end{equation}
In Eq. ({\ref{conditional.force}}) we have introduced the notation $\tilde{\sigma^x} = \sigma^z$, $\tilde{\sigma^y} = \sigma^y$, that is, $\tilde{\sigma^\alpha}$ is the component coupled to the motion in the direction $\alpha$.

\subsection{Effective spin--spin interactions}
Under certain conditions, a set of spins coupled to a common bath of vibrational modes is well described by an effective spin interacting Hamiltonian, something that is apparent if one makes use of the following canonical transformation \cite{Spin.Simulator,Wunderlich}:
\begin{eqnarray}
U = e^{-{\cal S}}, &&\hspace{0.2cm} {\cal S} = \sum_{\alpha,i,n} \eta^{\alpha}_{i,n} \left(1 + \tilde{\sigma}^{\alpha}_i \right)
 \left( a^{\dagger}_{\alpha,n} -  a_{\alpha,n} \right),
\nonumber \\
{\eta}^{\alpha}_{i,n} &=& F_{\alpha} \ \frac{{\cal M}^{\alpha}_{i,n}}{\hbar \omega_{\alpha,n}}
\sqrt{\frac{\hbar}{2 m \omega_{\alpha,n}}},
\label{canonical.transformation}
\end{eqnarray}
where $\eta^{\alpha}_{i,n}$ are the displacements of the modes in units of the ground state size. 

In the new basis the Hamiltonian (\ref{Hamiltonian.initial}) includes an effective spin-spin interaction:
\begin{equation}
e^{- {\cal S}} H e^{\cal S} = H_v + \frac{1}{2} \sum_{\alpha,i,j} J^{[\alpha]}_{i,j}
\tilde{\sigma}_i^{\alpha} \tilde{\sigma}_j^{\alpha} + \sum_{\alpha, i}
{B'}^{\alpha} \sigma_i^{\alpha} + H_E ,
\label{transformed.hamiltonian}
\end{equation}
where
\begin{equation}
- J^{[ \alpha ]}_{i,j} \! = \!
\sum_n \frac{F_{\alpha}^2}{m \omega^2_{\alpha,n}}
{\cal M}^{\alpha}_{i,n} {\cal M}^{\alpha}_{j,n} = 2 \sum_n \eta^{\alpha}_{i,n} \eta^{\alpha}_{j,n}  \hbar \omega_{\alpha,n} .
\label{effective.interaction.1}
\end{equation}
$H_E$ is a residual spin--phonon coupling, whose explicit form is given below.
From (\ref{effective.interaction.1}) and the definition of $\cal M^\alpha$ one can easily deduce that in the limit $\beta_\alpha \ll 1$:
\begin{equation}
J^{[\alpha]}_{i,j} \approx \frac{J^{[\alpha]}}{|{z'}^0_i - {z'}^0_j|^3} , \ \ J^{[\alpha]} = 2 \beta_\alpha \eta_\alpha^2 \hbar \omega_\alpha ,
\label{effective.interaction.simplified}
\end{equation}
where ${z'}^0_i$ are the ions' positions in units of the average distance, $d_0$. We have introduced $\eta$, which characterizes the displacement of the phonons due to the presence of the state--dependent force:
\begin{equation}
\eta_{\alpha} = F_{\alpha} \sqrt{\hbar/2 m \omega_\alpha}/\hbar \omega_\alpha .
\label{eta}
\end{equation}
Note that in (\ref{effective.interaction.simplified}) the effective magnetic fields receive, after the canonical transformation, a contribution from the pushing forces ${B'}^{\alpha} = B^{\alpha} + F_{\alpha}^2/(m \omega_{\alpha}^2)$. The extra term in ${B'}^{\alpha}$ does not depend on the ion's site, and thus can be considered as an overall correction to the global effective magnetic fields.

Radial modes in a chain of trapped ions allow us to implement two types of spin models. If $H_f$ acts only on one of the directions of motion, say ${\bf x}$, we get an Ising spin--spin interaction:
\begin{equation}
H^{Ising}_S = \frac{1}{2} \sum_{i,j} J^{[x]}_{i,j} \sigma_i^z \sigma_j^z + \sum_i B^x \sigma_i^x .
\label{effective.ising}
\end{equation}
The residual spin--phonon couplings are given, to lowest order in $\eta$ by:
\begin{eqnarray}
H^{Ising}_E &=&  B_x \sum_{i,n}  -i \sigma_i^y \eta^{x}_{i,n} (a^{\dagger}_{x,n} \! - \! a_{x,n}).
\label{error.ising}
\end{eqnarray}
On the other hand, if we apply two conditional forces in both radial directions, we get a Hamiltonian which couples two components of the effective spins.
\begin{equation}
H^{XY}_S = \frac{1}{2} \sum_{i,j} (J^{[x]}_{i,j} \sigma_i^z \sigma_j^z + J^{[y]}_{i,j} \sigma_i^y \sigma_j^y) + \sum_i B^x \sigma_i^x .
\label{effective.XY}
\end{equation}
A magnetic field $B^x$ is included such that, in a rotated basis $\sigma^z \rightarrow \sigma^x$, $\sigma^x \rightarrow -\sigma^z$, and (\ref{effective.XY}) corresponds to the XY model in its usual basis. The residual spin--phonon couplings are given, to lowest order in $\eta_\alpha$ by:
\begin{eqnarray}
H^{XY}_E &=&  - \frac{1}{2} \underset{i,n,m}{
\underset{\alpha,\alpha' = z,y}{\sum} }
\eta^{\alpha}_{i,n} \eta^{\alpha'}_{i,m} \hbar \omega_{\alpha, n}
\nonumber \\
& & ( a^{\dagger}_{\alpha,n} + a_{\alpha,n} ) ( a^{\dagger}_{\alpha',m} - a_{\alpha',m} )
\left[ \tilde{\sigma}_i^{\alpha}, \tilde{\sigma}_i^{\alpha'}  \right] .
\label{error.XY}
\end{eqnarray}
$H^{XY}_E$ accounts for the interference between the two different conditional forces, which can be avoided by choosing different values for the two radial trapping frequencies, as we explain in the following subsection.

\subsection{Decoherence induced by vibrational modes}

The residual spin--phonon couplings (\ref{error.ising}, \ref{error.XY}) are a source of decoherence which entangle internal states with phonons. Besides that, internal states are prepared and measured in a different basis than the one corresponding to effective interacting spins. Both effects have to be evaluated to study how the quantum dynamics of internal states departs from the ideal quantum spin Hamiltonian. The problem of solving the quantum dynamics of a system of interacting spins coupled to phonons is indeed quite complicated. We are interested here in estimating the error induced by these couplings, so that we will make use of perturbation theory and several approximations.

Let us consider that the system is initially in a product state of a thermal phonon density matrix, $\rho_{ph}$, and a given internal pure state, $\rho_i = | \psi_i \rangle \langle \psi_i |$. $| \psi_i \rangle$ evolves to $| \psi_f \rangle$ under the spin Hamiltonian $H_S$, thus, $|\psi_f \rangle$ represents the ideal simulated spin state. We define the fidelity in the quantum simulation by the overlap between the state of the system after the evolution with the whole Hamiltonian, and $| \psi_f \rangle$:
\begin{equation}
{\cal F}(\psi_i) = 
\langle \psi_f | \Tr_{ph} \{ e^{-i H t/\hbar} \ \rho_i \otimes \rho_{ph} \ e^{i H t/\hbar}  \} | \psi_f \rangle. 
\label{fidelity1}
\end{equation}
Having in mind perturbative calculations, let us express the fidelity in the following way:
\begin{eqnarray}
&& {\cal F}(\psi_i) =  \label{fidelity}
\\
&& \langle \psi_i | \Tr_{ph} \{ 
e^{{\cal S}(t)} U(t)  e^{-{{\cal S}(0)}}
 \rho_i \! \otimes \! \rho_{ph}
 e^{{\cal S}(0)} U(t)^{\dagger} e^{-{{\cal S}(t)}} \} 
 | \psi_i \rangle, \nonumber
\end{eqnarray}
where $U(t) \equiv e^{i H_0 t/\hbar} e^{-i(H_0 + H_E)t/\hbar}$, is the evolution operator in the interaction representation with respect to $H_E$. $H_0 = H_v + H_S$ is the Hamiltonian without residual spin--phonon coupling. In Eq. (\ref{fidelity}), as well as in the right--hand side of the equations below, all the operators evolve with $H_0$.

We are particularly interested in the very practical question of how do measurements of internal state observables departure from the averages in the simulated quantum spin models. Let us consider a few--site spin operator, $O$, and define the error in its average, $E_O = \langle O(t) \rangle - \langle O(t) \rangle_0$, with:
\begin{eqnarray}
&& \langle O(t) \rangle =  \nonumber \\
&& \ \ \textmd{Tr} \{ O(t) e^{S(t)} U(t) e^{-S(0)}
\rho_i \! \otimes \! \rho_{ph} e^{S(0)} U(t)^{\dagger} e^{-S(t)} \} , \nonumber \\
&& \langle O(t) \rangle_0 = \langle \Psi_f | O |\Psi_f \rangle .
\label{error.observables}
\end{eqnarray}

Equations (\ref{fidelity}) and (\ref{error.observables}) provide us with a well suited starting point for calculating the effects of the residual spin--phonon coupling in a series in $\eta_\alpha$. We show below that these terms can result in negligible contributions with the right choice of parameters. Due to the fact that lowest order spin--phonon couplings are different in each quantum spin model, we consider the two cases separately.

\subsubsection{Ising Model}
The spin--phonon couplings in (\ref{error.ising}) are proportional to the transverse magnetic field, $B^x$. In the following we ignore the index $\alpha$ in the vibrational modes, and assume that it corresponds to one of the radial directions. In the most interesting regimes $B^x \approx J$, thus we estimate $B^x \approx \eta^2 \omega$. All terms in $H_E^{Ising}$ are non--resonant, such that, if we ignore the contributions from $\cal S$ in (\ref{fidelity}, \ref{error.observables}), the only allowed transitions are virtual with probability $E \approx J^2 \eta^2/\omega^2 = {\cal O}(\eta^6)$.

The most important contribution to the error is thus the one that comes from the canonical transformation only, which can be estimated by setting $U(t) = {\bf 1}$, expanding $e^{S(t)}$ in (\ref{fidelity}), and keeping terms up to order $\eta^2$. In the stiff limit, that is $\beta \ll 1$, we make the additional approximation of considering vibrational modes in (\ref{fidelity}) as localized phonons, and get the following expression for the error to lowest order in $\eta$:
\begin{eqnarray}
E &=& 1 - {\cal F} = E^{[0,0]} + E^{[t,t]} -  E^{[0,t]}
- E^{[t,0]}, 
\nonumber \\
E^{[t_1,t_2]} &=&  \eta^2 \sum_{j} ( \bar{n} e^{i \omega (t_1-t_2)} + (\bar{n}+1)e^{-i \omega (t_1-t_2)} ) \times \nonumber \\
&& \hspace{.5cm} \left( \langle \sigma^z_j(t_1) \sigma^z_j(t_2) \rangle -
          \langle \sigma^z_j(t_1) \rangle \langle \sigma^z_j(t_2) \rangle \right) ,
\label{error.ising.2}
\end{eqnarray}
where $\bar{n}$ is the mean radial phonon number. The error is of order $N \eta^2$, and is proportional to the fluctuations in $\sigma^z_j$. Note that equal--time averages give the dominant contribution in (\ref{error.ising.2}) because the two--time spin averages in $E^{[0,t]}$ can be neglected at long enough times.

The scaling of $E$ with $N$ results from the fact that ${\cal F}$ in (\ref{fidelity}) represents the overlap between the internal state of the ion chain and the effective spin state. However, in practice, local observables like single spin averages or two--site correlation functions are measured. We can show that in this case the error does not increase with $N$. Let us consider again the effect of the canonical transformation in Eq. (\ref{error.observables}). We get the following error in the measurement of the observable:
\begin{eqnarray}
E_{O} &=& E_O^{[0,0]} + E_O^{[t,t]} -  2 E_O^{[0,t]} ,
\nonumber \\
E_O^{[t_1,t_2]} &=&
\frac{1}{2} \langle [[O(t),S(t_1)],S(t_2)] \rangle .
\label{error.ising.operator.2}
\end{eqnarray}
An explicit expression can be derived from (\ref{error.ising.operator.2}) which is not very enlightening.
However we note again that equal time correlations in (\ref{error.ising.operator.2}) are the dominant contribution, such that:
\begin{eqnarray}
E_O \approx \frac{1}{2} \eta^2
\sum_j (2 \bar{n} + 1) \langle [[O(t),\sigma_j^z(t)],\sigma_j^z(t)] \rangle ,
\label{error.ising.operator.3}
\end{eqnarray}
where we have also approximated vibrational modes by localized phonons. It is clear from Eq. (\ref{error.ising.operator.3}) that if $O$ is an M--site observable, there are only a maximum of $M$ non-vanishing commutators and thus $E \approx M \eta^2 \omega$. The most meaningful physical quantities in the study of quantum criticality are indeed one--site (mean values) or two--site (correlation functions) averages, and (\ref{error.ising.operator.3}) implies that these ones can be studied with an error that is independent on the number of ions.
\subsubsection{XY Model}

In this case special care has to be paid to the effect of the
residual spin--phonon couplings. $H_E^{XY}$ in (\ref{error.XY}) is
of order $\eta_\alpha^2 \omega_\alpha$, that is, of the same order as $J^{[\alpha]}_{i,j}$
itself. On the other hand, if $\omega_x = \omega_y$, then $H_E$
contains resonant terms that couple vibrational modes in different
transverse directions. Under these conditions the effect of $H_E$
is comparable to that of the spin--spin interaction and the
quantum simulation is ruined.

A way out of this problem is to tune $\omega_x \neq \omega_y$, with $\omega_x - \omega_y$ of the order of $\omega_x$, $\omega_y$. In this case, there are no resonant terms in (\ref{error.XY}), and the interference between standing--waves in different radial directions is suppressed. The error is then of the order of $(\eta^2 \omega_{x,y})^2 / (\omega_x - \omega_y) \approx (\eta_x^4,\eta_y^4)$. Indeed, Paul traps are usually designed such that the radial frequencies are different, with parameters that fulfill the conditions for the rotating terms in $H_E^{XY}$ to be neglected \cite{Leibfried.review}. Under these conditions, the lowest order contribution to the error is, again, the one that comes from the change of basis ($E \propto \eta_\alpha^2 \omega_{\alpha}$).

\section{Effective Ising model}
\label{effective.ising.model}
We have shown that antiferromagnetic long--range Ising models can be realized in experiments with ion traps. Hamiltonian (\ref{effective.ising}) is exactly solvable in the case of interaction between nearest--neighbors (NN) \cite{Sachdev,Ising.book}. In this case, the sign of the interaction is not relevant at all, because the transformation:
\begin{equation}
U = \prod_{j \ \textmd{odd}} \left( \sigma^x_{j} \right), \ \
\sigma^z_j \rightarrow U \sigma^z_j U^{-1} = (-1)^{j} \sigma^z_j ,
\label{AF.F.mapping}
\end{equation}
maps the ferromagnetic into the antiferromagnetic model.

The exact solution of the NN--Ising model shows that there exists a quantum phase transition at $B^x_c = J$ \cite{Sachdev} between a paramagnetic state ($B^x > J$), and an antiferromagnetic phase ($B^x < J$) characterized by the N\'eel order parameter:
\begin{equation}
O_N \equiv \frac{1}{N} \sum_j (-1)^j \sigma^z_j .
\label{Neel.OP}
\end{equation}
We expect that the properties of the effective spin model in ion traps (\ref{effective.ising}) are similar to those of the NN--Ising model, due to the fast decay of interactions $J_{i,j} \propto 1/|i-j|^3$; in particular, we expect a quantum phase transition at a given critical value of the longitudinal magnetic field, $B^x_c$. In the following we present numerical calculations to describe quantitatively the quantum phases of Hamiltonian $(\ref{effective.ising})$, and show that critical properties are very similar to those of the NN--Ising model. Indeed, renormalization group arguments can be used to show that the $1/r^3$ Ising model belongs to the short--range Ising universality class \cite{Dutta}. On the other hand, the long range character of the interactions turns out to induce intriguing effects in the spin quantum correlations which are explained below by means of a spin--wave model.

This numerical problem is handled with the Density Matrix Renormalization Group (DMRG) method \cite{White},
which is a quasi--exact numerical method for the study of ground states of
interacting quantum systems in 1D.
The fact that we have further than nearest--neighbor terms increases the complexity of
the algorithm by a factor of the number of sites, $N$. We keep $m$ = 128 eigenstates of the reduced density matrix at each step in the
DMRG algorithm and test the accuracy of our calculation by comparing its results with the exact solution in the NN--Ising model at the critical point, where DMRG works worst.
We have found that the relative error of the ground
state energy in this case is limited by the machine accuracy $\delta E \sim 10^{-15}$.
One can expect the same accuracy in calculations with the ion trap Ising models presented below because correlations are similar in
both cases.

Our numerical calculations describe two types of quantum Ising models:
\begin{itemize}
\item[(i)] {\it $1/r^3$--Ising interaction.}

This case corresponds to a linear array of microtraps, where equilibrium positions of the ions are approximately constant and fixed by the position of the microtrap.

\item[(ii)] {\it Linear trap--Ising model.}

If the trap is in the stiff limit, then effective spin--spin interactions decay like $1/|z^0_i - z^0_j|^3$, but distances between ions in the Coulomb crystal depend on the position. Thus, we get an effective inhomogenous quantum Ising model with interaction strength $J_{i,j}$ which depends on the position (see Fig. \ref{magnetization}). The ground state shows the coexistence of different phases in different locations of the ion trap.
\end{itemize}

The results presented in the following subsections where obtained with chains of $N$ = $100$ ions. $J_{i,j}$ in the linear trap--Ising model was calculated with $\beta_x = 10^{-2}$, however, our results do not change much with $\beta_x$, as long as one considers values within the stiff limit ($\beta_x < 0.1$). If we assume typical values $\omega_x$ $=$ 10 MHz, and $\eta^2 = 10^{-2}$, this would correspond to interaction strength $J_{i,i+1}$ $\leq$ 20 kHz.

\subsection{Effective magnetization}
\label{effective.magnetization}
In the following we study the effective magnetization and its fluctuations. We will be mainly interested in {\it (i)} whether global measurements are enough to characterize quantum phases, and {\it (ii)} which is the effect of inhomogeneity and finite size in linear trap models. All our results are presented in energy units $J_{0}$, where $J_0$ is the interaction strength between nearest--neighbors in the $1/r^3$--Ising model, or the averaged nearest--neighbor interaction $1/N \sum_{i} J_{i,i+1}$, in the case of the linear trap--Ising model.

\subsubsection{Transverse ($\langle \sigma^x \rangle$) magnetization.}

\begin{figure}[t]
  \center
  \includegraphics[width=\linewidth]{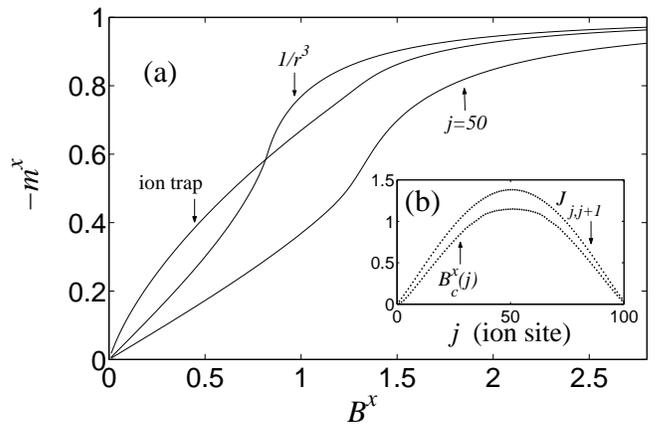}
  \caption{(a) Evolution of the averaged effective longitudinal magnetization in $1/r^3$ and linear trap Ising models ($N$=100 ions). We also plot $\langle \sigma^x_{j} \rangle$ in the case of the central ion ($j=50$) in a linear trap.
(b) Local strength of the Ising interaction in a linear ion trap, and local value of the critical field.}
  \label{magnetization}
\end{figure}

\begin{figure}[t]
\center
\includegraphics[width=1.6in,height=1.4in]{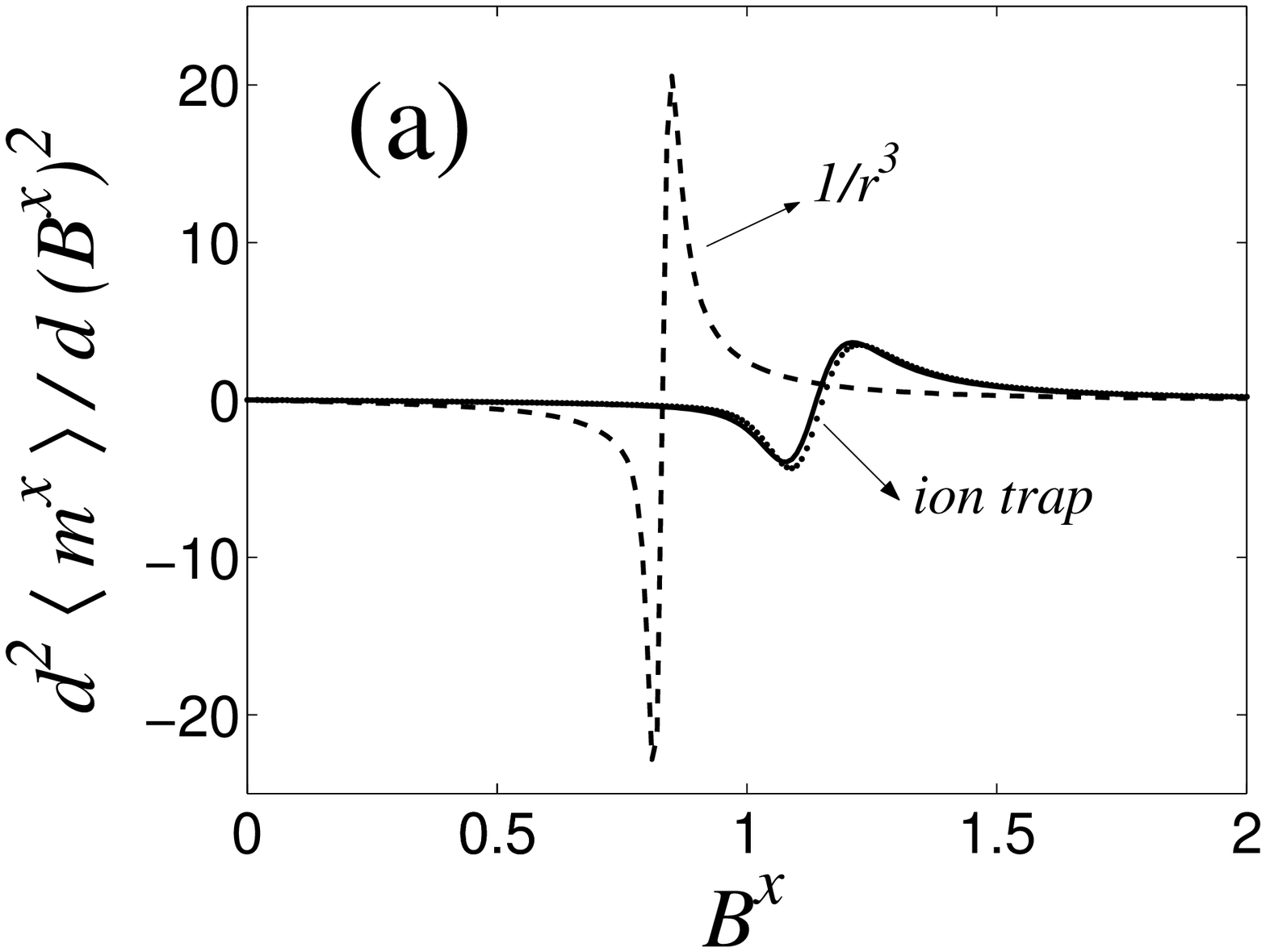}
\hspace{0.2cm}
\includegraphics[width=1.6in,height=1.4in]{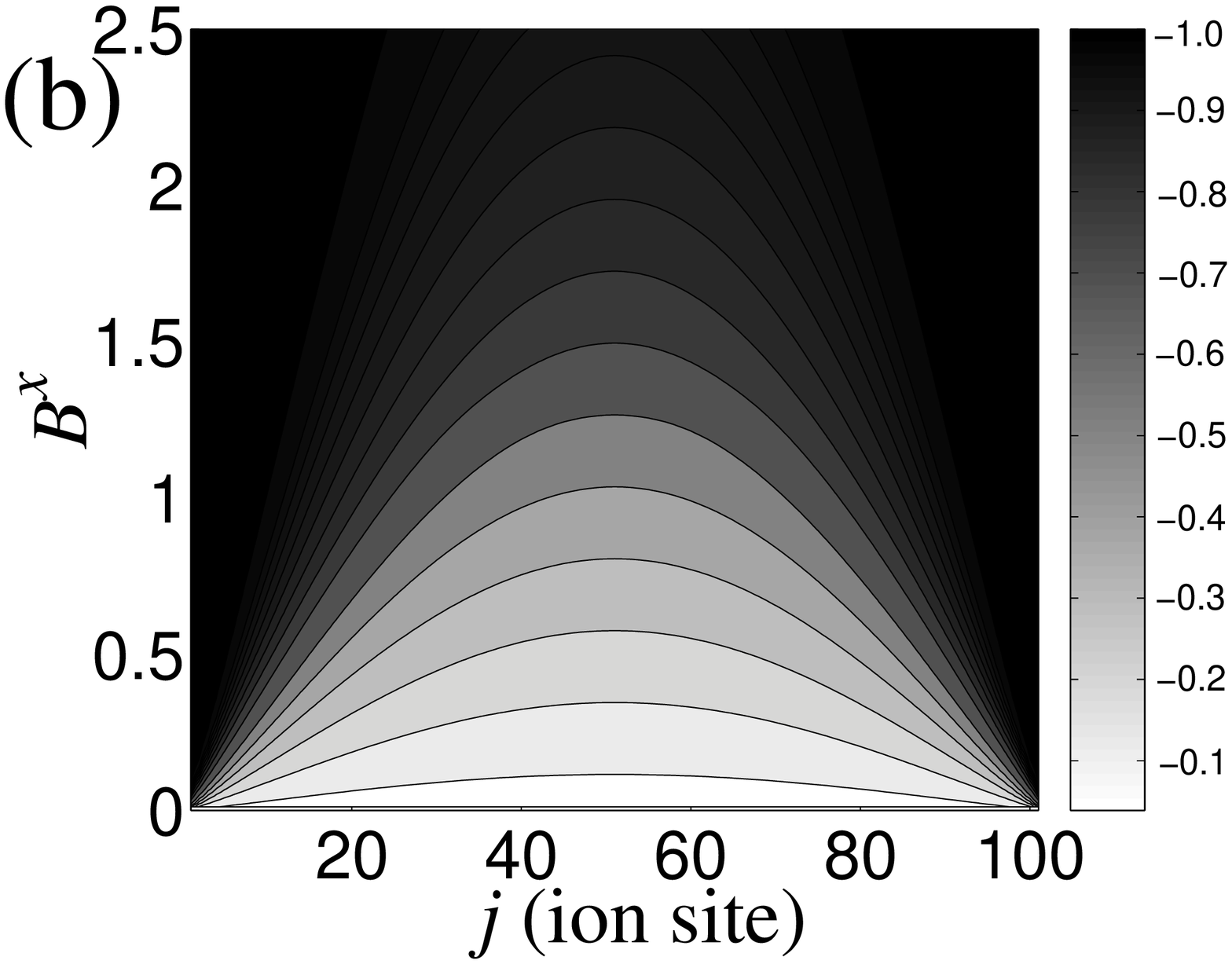}
\caption{(a) Second derivative of $m^x$ as a function of $B^x$ in ion trap Ising models ($N$ = 100 ions). 
Dashed line: $1/r^3$--Ising model. Solid line: average over the 20 central ions in a linear trap--Ising model. Dotted line: central ($j=50$) ion in a linear trap--Ising model (solid and dotted line are almost on top of each other). 
(b) Contour plot of $\langle \sigma^x_j  \rangle$ on the plane of the coordinate $j$ and the magnetic field $B^x$.}
\label{mxcontour}
\end{figure}

The mean magnetization, $m^x = (1/N) \langle \sigma^x_T
\rangle$ (where $\sigma^\alpha_T = \sum_j \sigma_j^\alpha$), can
be obtained from global measurements. 
In Fig. \ref{magnetization} (a) we show the evolution of $m_x$ with the magnetic field. The magnetization curve of the $1/r^3$--Ising model is similar to the NN case. The quantum phase transition results in a discontinuity in $d^2 m^x/ d (B^x)^2$ (see Fig. \ref{mxcontour}
(a)), something that allows us to locate the critical point $B^x_c(1/r^3) \approx 0.83$, which lies below the critical point in the NN- Ising model ($B^x_c(NN) = 1$). This effect can be explained in terms of frustration induced by terms $J_{i,j}$, with $(i-j)$ even, which reduces the stability of antiferromagnetic order. 

On the other hand, $m^x$ in the linear trap--Ising model departs from the homogeneous $1/r^3$--Ising case, due to the spatial variations in $J_{i,j}$ (see Fig. \ref{magnetization} (b)). The system shows the coexistence of different phases, depending on the local value of $J_{i,j}$, as shown in the evolution of the local magnetization $\langle \sigma^x_j \rangle$ with $B^x$ in Fig. \ref{mxcontour} (b). 
The local phase diagram at each ion $j$ shows a critical point $B^x_c(j) \approx 0.83 J_{j,j+1}$, which is determined by the local value of the interaction, but satisfies the same relation with $J_{i,i+1}$ found for the homogenous $1/r^3$--Ising model (Fig. \ref{magnetization} (b)).

The local phases in a linear ion trap Ising model can be studied with only limited local addressing of the ions. For example, by measuring the average magnetization, $m^x$, corresponding to the 20 central ions, we can observe the signature of the quantum phase transition, as shown in Fig. \ref{mxcontour} (a). Thus, individual ion addressing is not necessary for detecting the critical point in the quantum simulation. 

\subsubsection{Longitudinal ($\langle \sigma^z \rangle$) magnetization.}
\begin{figure}[t]
  \center
  \includegraphics[width=\linewidth]{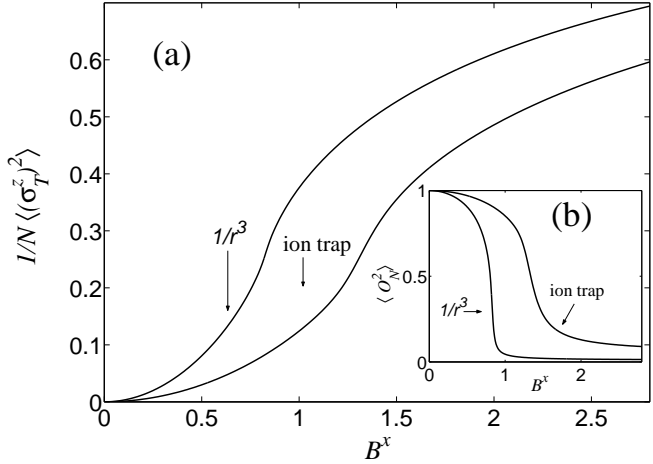}
  \caption{(a) Fluctuation of the longitudinal effective magnetization, and (b) N\'eel order parameter. In both plots we consider the $1/r^3$--Ising model, and the central region (20 ions) of the linear trap--Ising model.}
\label{fluctuation}
\end{figure}

The antiferromagnetic order parameter is given by the staggered
magnetization (\ref{Neel.OP}). In the thermodynamic limit,
spontaneous symmetry breaking results in a non--zero value of
$\langle O_N \rangle$. On the contrary, in finite systems, symmetry remains unbroken. In this case, it is convenient to study the squared antiferromagnetic order parameter $\langle O_N^2 \rangle$, which takes a value $\approx 1$ in the antiferromagnetic phase. In Fig. \ref{fluctuation}, we present the evolution of $\langle O_N^2 \rangle$ in the $1/r^3$--Ising case, as well as in the central region (20 ions) of a linear trap. 

For measuring $O_N$ it is necessary to address each ion individually. On the other hand, the fluctuation of the average longitudinal magnetization, $(1/N^2) \langle
(\sigma^z_T)^2\rangle$ is an interesting alternative which does
not require individual ion addressing. Antiferromagnetic order
can be detected by means of this global observable, because
longitudinal spin fluctuations are suppressed in the N\'eel
ordered state, as shown in the case of the $1/r^3$--Ising model
(Fig. \ref{fluctuation}). 
\subsection{Correlation Functions}
Correlation functions 
$C^{\alpha \alpha}_{i,j} \equiv \langle \sigma^\alpha_i \sigma^\alpha_j \rangle - \langle \sigma^\alpha_i \rangle \langle \sigma^\alpha_j \rangle$ characterize quantum correlations in the ground state of the effective spin system. Even when they play an important role in the description of strongly correlated systems, it is not possible to measure them directly in solid--state experimental setups \cite{Auerbach}. Trapped ions, on the contrary, offer us the possibility to measure directly equal--time correlation functions by means of a set of measurements on single ions. We show here that realizations of Ising models with trapped ions allow us to test directly properties of quantum critical systems such as the algebraic decay of correlations at a quantum phase transition, as well as remarkable new effects induced by long--range interaction terms.

We consider correlations of observables that are transverse to the order parameter, that is, $C^{xx}_{i,j}$ correlations in the antiferromagnetic phase and $C^{zz}_{i,j}$ correlations in the paramagnetic one, because they are the most meaningful in terms of the spin--wave picture to be introduced later. In Figs. \ref{IsingCorr} and \ref{IonCorr}, it is shown that quantum correlations present two regimes, depending on the distance between ions, $|i-j|$:

\subsubsection{Universality regime.}

\begin{figure}[t]
  \center
  \includegraphics[width=\linewidth]{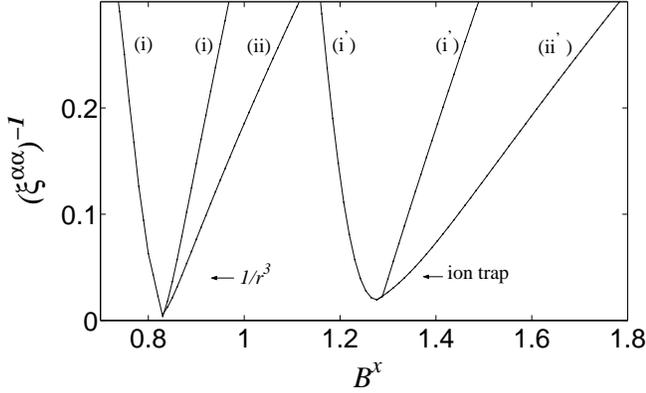}
  \caption{Correlation lengths $(\xi^{\alpha \alpha})^{-1}$ as a function of $B^x$ in both $1/r^3$ and linear ion trap models. 
(i), (i') curves correspond to $(\xi^{xx})^{-1}$, whereas (ii), (ii') correspond to $(\xi^{zz})^{-1})$. $\xi^{zz}$ is not shown in the antiferromagnetic phase because $C^{zz}_{i,j}$ tends to a constant value in the absence of symmetry breaking.}
\label{corrlength}
\end{figure}

At intermediate distances, correlation functions both in $1/r^3$, and in linear trap--Ising models, show critical properties which are similar to those of the NN--Ising model: 
$C^{\alpha \alpha}_{i,j} \propto e^{-|i-j|/\xi^{\alpha \alpha}}$ when $B^x \neq B^x_c$, whereas  $C^{\alpha \alpha}_{i,j} \propto |i-j|^{-p}$ at $B^x_c$, with $p=2$. On the other hand, correlation lengths $\xi^{\alpha \alpha}$ (see Fig. \ref{corrlength}) diverge near the critical point and show the dependence, $(\xi^{\alpha \alpha})^{-1} \propto |B^x - B^x_c|^\nu$, with $\nu \sim 1$. The critical properties of the $1/r^3$--Ising model are thus the same as those of the nearest--neighbor model \cite{Dutta}. 

We note that by measuring quantum correlations in the central region of the chain one can measure critical exponents even in the case of the linear--trap Ising model. Finite size effects are however more important in this case (see Fig. \ref{IonCorr} (b)).

\subsubsection{Long-range correlation mediated by the interaction.}

\begin{figure}[t]
  \center
  \includegraphics[width=\linewidth]{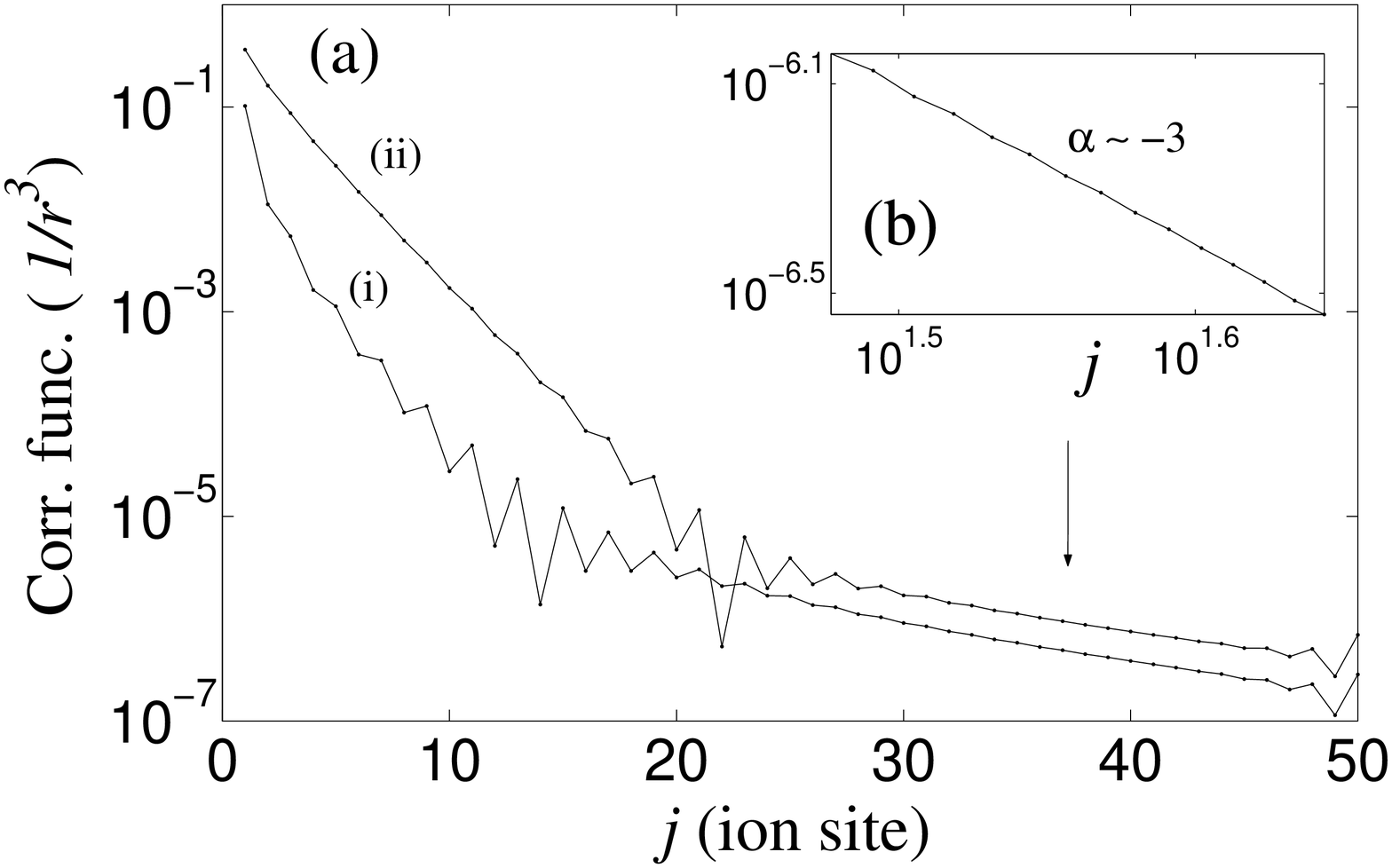}
  \caption{(a) Absolute value of correlations, $|C^{\alpha\alpha}_{j_0,j_0+j}|$, between the central ion ($j_0 = 50$) and the rest of a chain with $N=100$ ions, in the case of the $1/r^3$--Ising model.
(i) $C^{xx}_{j_0,j_0+j}$, 
$B^x=0.72 < B^x_c$; (ii)  $C^{zz}_{j_0,j_0+j}$, $B^x=1.32 > B^x_c$. (b) Zoom of the long--range tail of $C^{zz}$, which follows an algebraic decay with an exponent $\alpha = -3$.}
  \label{IsingCorr}
\end{figure}

\begin{figure}[t]
  \center
  \includegraphics[width=\linewidth]{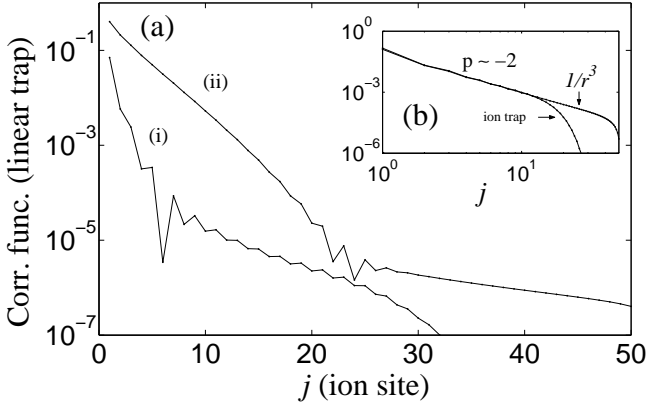}
  \caption{(a) Absolute value of correlations, 
$|C^{\alpha\alpha}_{j_0,j_0+j}|$, between the central ion ($j_0 = 50$) and the rest of a chain with $N=100$ ions, in the case of the linear trap--Ising model.
(i) $C^{xx}$, $B^x=0.89 < B^x_c$; (ii)  $C^{zz}$, $B^x= 1.72 > B^x_c$. (b) Plot of $C^{xx}_{j_0,j_0+j}$ for both $1/r^3$ and linear ion trap models, exactly at the critical point $B^x = B^{x}_c$.}
\label{IonCorr}
\end{figure}

A remarkable feature in Figs. \ref{IsingCorr} and \ref{IonCorr}, is that correlation functions decay like a power--law, $C^{\alpha \alpha}_{i,j} \approx 1/|i-j|^3$, at very long distances, so that properties of quantum correlations depart from the nearest--neighbor case (see Fig. \ref{IsingCorr} (b)).  

This effect can be qualitatively understood by considering that entanglement between distant ions can be created in two ways: (i) by nearest--neighbor terms in the Hamiltonian, in such a way that correlations present the same characteristics as those of the NN--Ising model (exponential decay), or (ii) directly by long--range terms in (\ref{effective.ising}) in such a way that they decay following the power law of the spin--spin interaction.
This hand--waving argument will be justified in the following section by means of a spin--wave formalism.


\subsection{Spin--wave picture}

Far away from the critical point, excitations over the ground state can be described as spin--waves satisfying a harmonic Hamiltonian. This picture allows us to understand the numerical results presented above; in particular correlations in the presence of power--law interactions. It also brings ion trap spin models in connection with recent results on quantum correlations and criticality in bosonic gaussian ground states \cite{Norbert.Michael}.

Spin--waves are defined by the Holstein--Primakoff (HP)
transformation \cite{Auerbach}, whose particular form depends on
the ground state. Thus we have two consider separately the two
limits corresponding to antiferromagnetic or paramagnetic phases.

\subsubsection{$B^x \gg J$}
In this phase, HP bosons describe spin--waves excited over the
paramagnetic ground state:
\begin{eqnarray}
(\sigma^z_j - i \sigma^y_j)/2 &=& b^\dagger_j (1-b_j^\dagger b_j)^{1/2} \ \approx b_j^\dagger , \nonumber \\
(\sigma^z_j + i \sigma^y_j)/2 &=& (1-b_j^\dagger b_j)^{1/2} b_j \approx b_j , \nonumber \\
\sigma^x_j &=& 2 b_j^\dagger b_j - 1 .
\label{HP.1}
\end{eqnarray}
The harmonic approximation is valid if $b_j^\dagger b_j \ll 1$. In this limit the effective spin Hamiltonian takes the following form:
\begin{equation}
H = \frac{1}{2} \sum_{j,l} J_{j,l}
(b^\dagger_j + b_j)(b^\dagger_l + b_l)
+  B^x  \sum_j (2 b_j^\dagger b_j - 1) .
\end{equation}
Let us write this Hamiltonian in terms of canonical operators:
\begin{eqnarray}
Q_l &=& \frac{1}{\sqrt{2}}(b^\dagger_l + b_l), \nonumber \\
P_l &=& \frac{i}{\sqrt{2}}(b^\dagger_l - b_l), \nonumber \\
H/|2 B^x| &=& \frac{1}{2} \sum_{j,l} K_{j,l} Q_j Q_l
+ \frac{1}{2} \sum_j P_j^2,
\label{Ising.HP}
\end{eqnarray}
with $K_{j,l} = J_{j,l}/|B^x| + \delta_{j,l}$.
In the following we consider the limit $N \rightarrow \infty$, so that we can get analytic results. In this limit, Hamiltonian (\ref{Ising.HP}) is diagonalized by plane--waves,
$\tilde{Q}_q = 1/\sqrt{N} \sum_j e^{i q j} Q_j$ and correlation functions are given by:
\begin{eqnarray}
C^{zz}_{0,j}
&=& 2 \langle Q_0 Q_j \rangle = \frac{2}{N} \sum_q e^{- i q j} \langle \tilde{Q}_q \tilde{Q}_{-q} \rangle
\nonumber \\
&=& \frac{1}{N} \sum_q e^{- i q j} \frac{1}{\Omega_q}.
\label{Czz.HP}
\end{eqnarray}
$\Omega_q$ is the spin--wave dispersion relation:
\begin{equation}
\Omega^2_q = \frac{1}{N} \sum_j K_{j,0} e^{i q j} .
\label{sw.dispersion}
\end{equation}
It is illuminating to write the correlation function in the following way:
\begin{eqnarray}
C^{zz}_{0,j} = \langle \sigma^z_0 \sigma^z_j \rangle &=& \frac{1}{N} \sum_q e^{- i q j} \Omega^2_q \frac{1}{\Omega^3_q} =
\sum_{l}  K_{j,l} A^z_l = \nonumber \\
& & A^z_j + \sum_l \frac{J_{j,l}}{B^x} A^z_l,
\label{Czz.HP.guays}
\end{eqnarray}
with:
\begin{eqnarray}
A^z_j &=& \frac{1}{N} \sum_{q} \frac{e^{i q j}}{\Omega_q^3} \approx \frac{1}{2 \pi} \int_{-\pi}^{\pi} dq \frac{e^{i q j}}{\Omega_q^3} .
\label{A.definition}
\end{eqnarray}
The function $A^z_j \rightarrow 0$ as $j \rightarrow \infty$, such that for long distances, Eq. (\ref{Czz.HP.guays}) implies that correlations decay following the spin--spin interaction power law. Indeed, the decay of correlations with the same power--law than the interaction term, has been recently shown to be a general property of ground states of harmonic lattices with long--range interactions \cite{Norbert.Michael}.

%
\subsubsection{$B^x \ll J$}
In this limit the ground state is close to the antiferromagnetic (N\'eel state), so that we use the following HP transformation:
\begin{eqnarray}
(-1)^j \sigma^z_j &=& 2 b_j^\dagger b_j - 1, \nonumber  \\
(\sigma^x_j + i (-1)^j \sigma^y_j)/2 &=& b_j^\dagger (1-b_j^\dagger b_j)^{1/2} \ \approx b_j^\dagger , \nonumber \\
(\sigma^x_j - i (-1)^j \sigma^y_j)/2 &=& (1-b_j^\dagger b_j)^{1/2} b_j \approx b_j .
\label{HP.2}
\end{eqnarray}
Signs, $(-1)^j$, have to be added to choose the N\'eel state as the ground state of the HP oscillators. The Hamiltonian is:
\begin{eqnarray}
H &=& \frac{1}{2} \sum_{j,l} (-1)^{j-l} J_{j,l} \left( 2 b_j^\dagger b_j - 1 \right) \left( 2 b_l^\dagger b_l - 1 \right) + \nonumber \\
& & B^x \sum_j \left( b^\dagger_j + b_j \right)  .
\label{Ising.HP.2}
\end{eqnarray}
If we neglect interactions between bosons, then we get a set of non--coupled harmonic oscillators. However, non--quadratic terms in (\ref{Ising.HP.2}) induce correlations due to the presence of the transverse field $B^x$. To see this, we solve first the quadratic part of the bosonic Hamiltonian:
\begin{eqnarray}
H_0 = \sum_j 2 \tilde{J} b_j^\dagger b_j + B^x \sum_j \left( b^\dagger_j + b_j \right) ,
\label{Ising.HP.2.0}
\end{eqnarray}
where $\tilde{J} = - \sum_j (-1)^{j-l} J_{j,l}$, that is, the mean longitudinal magnetic field. Hamiltonian (\ref{Ising.HP.2.0}) is
solved by displacing the HP bosons, which corresponds to consider
the mean-field ground state as a reference state for the HP
transformation:
\begin{equation}
b_j \rightarrow b_j - B^x/(2 \tilde{J}).
\label{HP.displacement}
\end{equation}
Up to quadratic terms in the displaced HP bosons, we get the following Hamiltonian:
\begin{eqnarray}
&& \hspace{-0.5cm} H = 2 \tilde{J} \sum_j b^{\dagger}_j b_j + \nonumber \\
&& \left( \frac{B^x}{2\tilde{J}} \right)^2 \sum_{j,l} (-1)^{(j-l)}
J_{j,l} \left( b^{\dagger}_j + b_j \right) \left( b^{\dagger}_l +
b_l \right).
\end{eqnarray}
By following the same steps as in the previous case, we show that:
\begin{eqnarray}
C^{xx}_{0j} = A^x_j + 
\sum_l (-1)^{j-l} \frac{B^x J_{j,l}}{\tilde{J}^2} A^x_l ,
\label{Cxx.HP.guays}
\end{eqnarray}
with $A^x_l$ given by the same function (\ref{A.definition}) of the spin--wave energies that diagonalize (\ref{Ising.HP.2.0}). Thus, to lowest order in $B^x/\tilde{J}$, x--x correlations behave in a similar way as z--z correlations in the antiferromagnetic phase, with the only difference being the alternation in the sign.

\begin{figure}
\center
  \includegraphics[width=\linewidth]{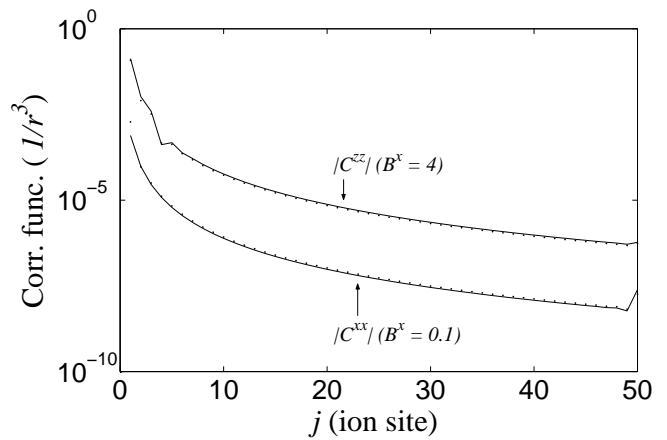}
  \caption{(a) Absolute value of correlations $|C^{\alpha\alpha}_{j_0,j_0+j}|$ ($j_0 = 50$, $N=100$ ions) in the case of the $1/r^3$--Ising model. Points: Numerical results from the DMRG calculations. Solid line: calculation with the HP transformation.}
\label{CorrFit}
\end{figure}

In Fig. \ref{CorrFit} we show the comparison between results obtained by means of the spin-wave picture and DMRG calculations in both antiferromagnetic and paramagnetic phases. We have checked that the agreement is good far away from the critical point, and gets worse when one approaches it because the assumptions behind the HP approximation are no longer valid. 

\section{Effective XY model}
We consider the isotropic case of Hamiltonian (\ref{effective.XY}), $J^{[x]}_{i,j}=J^{[y]}_{i,j} > 0$. Let us work in the rotated basis defined below Eq. (\ref{effective.XY}), with $B^z \rightarrow - B^z$:
\begin{equation}
H^{XY}_S = \frac{1}{2} \sum_{i,j} J_{i,j} ( \sigma_i^x \sigma_j^x
+ \sigma_i^y \sigma_j^y) + \sum_i B^z \sigma_i^z ,
\label{effective.XY.2}
\end{equation}
but will keep in mind that in experiments effective spin observables must be measured in the original basis.  Note also that $[H,\sum_j \sigma^z_j]=0$ and hence $\sigma^z_T$ is a conserved quantity.

The nearest-neighbor (NN) XY model can be exactly solved by a
Jordan--Wigner mapping to free fermions \cite{Lieb}. Note that here we consider the antiferromagnetic model, which can be mapped onto the ferromagnetic one by (\ref{AF.F.mapping}). The whole region $|B^z|/J<1$ is critical and $C^{xx}_{i,j}$ follow a power law with critical exponent $\alpha=1/2$. $\langle \sigma_x
\rangle$ = $\langle \sigma_y \rangle$ = $0$, whereas $\langle
\sigma_z \rangle$ grows as a function of $B^z$ up to the
non--analytical point $|B^z_c|/J=1$. One can expect that
properties of XY models in trapped ions are similar to those of the NN case.

In this section we study the two cases of $1/r^3$-XY interactions
and linear trap-Ising interactions (see the beginning of section
(\ref{effective.ising.model}) for a motivation of this distinction) by means of DMRG. $m=128$ eigenstates of the reduced density matrix are kept, and comparison
with the exact solution allows us to estimate a relative error in
the calculation of the energy, $\delta E \sim 10^{-13}$. We consider ion chains with $N=50$ ions, and the same parameters for the ion linear trap, and energy units, considered in the previous section.

\subsection{Effective magnetization}
\begin{figure}
  \center
  \includegraphics[width=\linewidth]{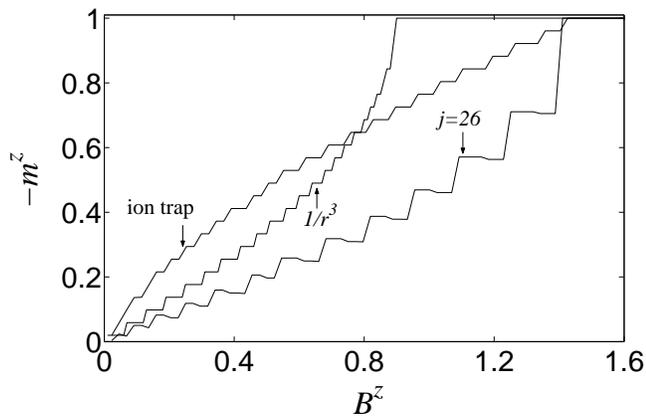}
  \caption{Phase diagram of $1/r^3$ and linear trap--XY models. We plot the averaged global magnetization, as well as the evolution of $\langle \sigma^z_j \rangle$ at the center of a linear trap ($N = 50$ ions).}
  \label{XYmag}
\end{figure}

The most interesting single spin observable is the magnetization in the ${\bf z}$ direction. Here we will follow the same lines and definitions explained in subsection \ref{effective.magnetization}. In Fig. \ref{XYmag} we plot the evolution of $m^z$ as a function
of $B^z$. The steps in the curve are due to the finite size of the ion chains. The magnetization curve of the $1/r^3$--XY model follows the same relation as in the nearest--neighbor model, $m_z \propto (B^z)^2$. We find again the same effect that in the Ising model, that is, the critical point is shifted $B^z_c(1/r^3) (\approx 0.9)$ $<$ $B^z_c(NN) (=1)$ due to frustration induced
by long-range interaction terms. 

\begin{figure}
  \center
  \includegraphics[width=1.65in]{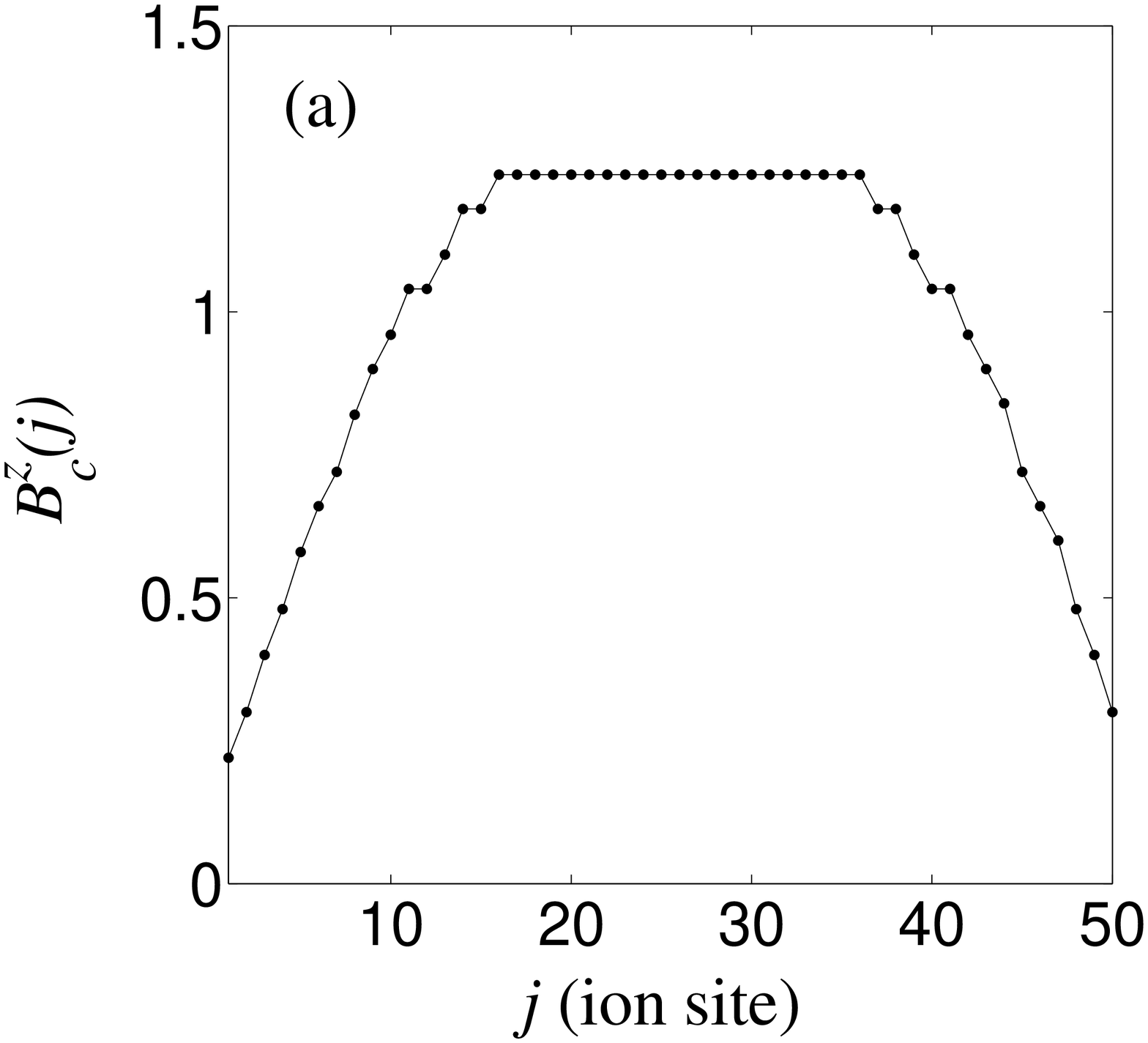}
  \includegraphics[width=1.65in]{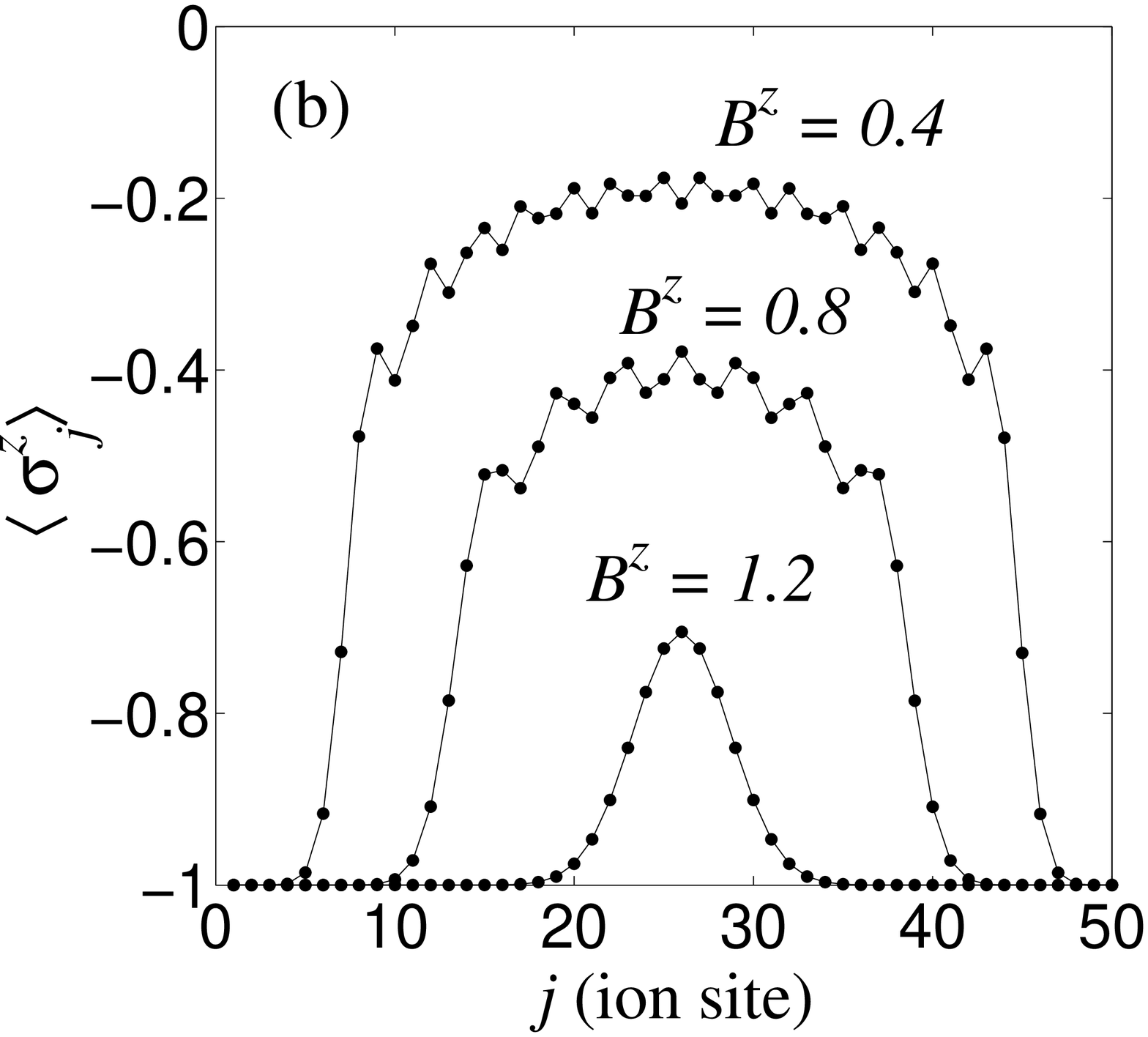}
  \caption{(a) Local critical field calculated from the evolution of $\langle \sigma^z_j \rangle$ in the linear trap--XY model. (b) Spatial dependence of $\langle \sigma^z_j \rangle$ in  a linear--trap XY model with different values of $B^z.$}
\label{XYlocal}
\end{figure}

On the other hand, in the linear trap-XY model, $m^z$ departs from the homogenous $1/r^3$ case, due again to the variations of the interaction strength along the ion trap. As we did in the case of the Ising model, we get a local phase diagram by plotting the evolution of the single site magnetization as a function of $B^z$. An interpretation of our results in terms of local quantum phases governed by the local value of $J_{i,j}$ is, however, not justified in this case, because of the existence of long--range correlations in the critical region of the XY model. This fact is shown in the dependence of the local critical field which does not match the spatial profile of the spin--spin interaction (Fig. \ref{XYlocal} (a)). In both $1/r^3$ and linear trap--XY models, $B^z_c$ can be calculated exactly \cite{Pazmandi}, and the result agrees with our numerical calculation.

A better picture of the linear trap-XY model can be obtained in terms of Jordan--Wigner fermions \cite{Ising.book}. If one neglects long-range terms, which lead to fermion-fermion interactions, then $\sigma^z_j$ corresponds to the local density of Jordan--Wigner fermions, and the evolution of $\sigma^z_j$ shows the emptying of fermionic levels as $- B^z$ (which plays the role of a chemical potential) is decreased (Fig. \ref{XYlocal} (b)). 

\subsection{Correlation functions}

\begin{figure}
  \center
  \includegraphics[width=\linewidth]{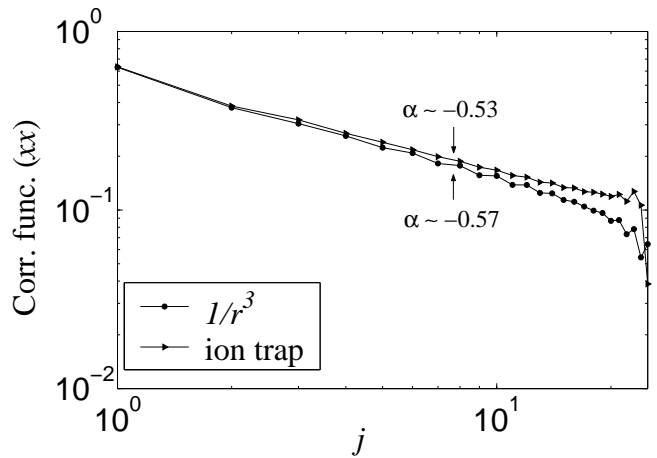}
  \caption{Correlation $C^{xx}_{j_0,j_0+j}$ between the central ion $j_0=26$ and the rest of the chain in the critical region as a function of ion separation $j$. $C^{xx}_{j_0,j_0+j}$ is well fitted by a power--lay decay with the exponents $\alpha$ shown in the figure.}
\label{XYcorr}
\end{figure}

Our DMRG calculations show that the phase $B^z < B^z_c$ is also critical in both $1/r^3$ and linear trap--XY models, as evidenced in the algebraic decay of $C^{xx}_{i,j} \propto 1/|i-j|^\alpha$. In this case, contrary to the Ising model, critical exponents are slightly different than in the NN--XY model, in which $\alpha = 1/2$ (see Fig. \ref{XYcorr}). Note that experiments with linear ion traps can detect the algebraic decay of correlation functions in XY models, even in the presence of the finite size effects induced by variations of $J_{i,j}$. 
\section{Conclusions}
In this work we have discussed in detail a recent
proposal \cite{Spin.Simulator} for the realization of quantum spin systems with trapped ions under the action of off-resonant
standing waves. Under certain conditions the coupling between internal states and
vibrational modes can be written as an effective spin-spin
interacting Hamiltonian with a residual spin-phonon coupling. 
In this way, the physics of quantum criticality can be accessed in experiments with ion traps. Our numerical calculations show that: 

\begin{itemize}
\item[(i)] 
In the homogeneous $1/r^3$-Ising model, which can be realized with ion microtraps there is a quantum phase transition with critical field $B^x_c\approx 0.83J$, and same critical properties as the nearest--neighbor Ising model.

\item[(ii)] In linear ion traps, due to the non-constant separation of the ions, the spin--spin interaction is inhomogeneous, which leads to the coexistence of different quantum phases. However, critical properties can be accessed by measuring each region in the trap separately, which only requires partial ion local measurements (10-20 ions).

\item[(iii)] Ion trap Ising models show long--range quantum correlations that are not present in nearest--neighbor models and can be explained by means of a spin--wave theory.

\item[(iv)] In the case of the $1/r^3$-XY model the critical field is shifted to a value $B^x_c \approx 0.9$. 
The quantum phase diagram can be determined by measuring the effective longitudinal magnetization, $m^z$. Besides that, experiments with trapped ions can access the algebraic decay of correlation functions in the critical phase of XY models.
\end{itemize}

\acknowledgements We thank Michael Wolf and Norbert Schuch for interesting discussions. Work supported by MEIF-CT-2004-010350, Deutscher Akademischer Austausch Dienst, CONQUEST and SCALA.

\newpage

\end{document}